\documentclass[aps,prr,twocolumn,amsmath,amssymb,superscriptaddress,longbibliography]{revtex4-1}
\usepackage[english]{babel}
\usepackage{xcolor}
\usepackage{amssymb}
\usepackage{dcolumn}
\usepackage{bm}
\usepackage{graphicx}
\usepackage{amsmath}
\usepackage{braket}

\usepackage{graphicx}        % standard LaTeX graphics tool

\graphicspath{{pict/}{}}

\usepackage[normalem]{ulem}

\usepackage{dcolumn}%%%%%new
\usepackage{bm}
\usepackage[pdfstartview=FitH, CJKbookmarks=true, bookmarksnumbered=true, bookmarksopen=true, colorlinks=true, pdfborder=001, citecolor=blue, linkcolor=blue, urlcolor=blue, linktocpage=true] {hyperref}

\begin{document}

\title{Spin-Exchange-Controlled Nonclassical Light in Cavity Electromagnetically Induced Transparency}

\author{Jing Tang}
\email{jingtang@gdut.edu.cn}
\affiliation{School of Physics and Optoelectronic Engineering, Guangdong University of Technology, Guangzhou 510006, China}
\affiliation{Guangdong Provincial Key Laboratory of Sensing Physics and System Integration Applications, Guangdong University of Technology, Guangzhou, 510006, China}

\author{Junyu Su}
\affiliation{School of Physics and Optoelectronic Engineering, Guangdong University of Technology, Guangzhou 510006, China}

\author{Jie Chen}
\affiliation{School of Physics and Optoelectronic Engineering, Guangdong University of Technology, Guangzhou 510006, China}

\author{Ziyang Li}
\affiliation{School of Physics and Optoelectronic Engineering, Guangdong University of Technology, Guangzhou 510006, China}

\author{Yuangang Deng}
\email{dengyg3@mail.sysu.edu.cn}
\affiliation{Guangdong Provincial Key Laboratory of Quantum Metrology and Sensing $\&$ School of Physics and Astronomy, Sun Yat-Sen University (Zhuhai Campus), Zhuhai 519082, China}

\date{\today}

\begin{abstract}
Controlling nonclassical light while maintaining a finite cavity response requires selective suppression of unwanted multiphoton excitation. We investigate a ring-cavity system containing two three-level emitters coupled by a coherent spin-exchange interaction, where electromagnetically induced transparency (EIT) provides resonant access to a dark-state polariton. The exchange interaction shifts the polariton resonance while modifying the dressed two-photon excitation pathways. Strong photon antibunching emerges when the one-polariton resonance nearly coincides with a destructive-interference minimum of the second-photon transition amplitude. By jointly tuning the exchange interaction and control-field strength, pronounced photon blockade with a finite intracavity population can be sustained over an extended interaction range. We further show that the relative cavity-coupling phase redistributes optical coupling between symmetric and antisymmetric collective channels with opposite exchange shifts, allowing the photon statistics to switch between antibunching and bunching at fixed exchange strength. This mechanism provides a possible route to tunable nonclassical-light sources based on coherent control of collective excitation pathways.
\end{abstract}

\maketitle

\section{Introduction}
Generating nonclassical light at the single-photon level is a central goal of quantum optics, with applications in quantum communication, photonic quantum information processing, and quantum networks~\cite{Bennett00,Kimble08,Brien09,RevModPhys.79.135,RevModPhys.90.035005}. Photon blockade provides a route to antibunched emission by suppressing the excitation of successive photons~\cite{Imamoglu97,Birnbaum2005,Faraon08,Reinhard2012}. Conventional photon blockade exploits spectral anharmonicity to separate one- and two-photon resonances, with additional control provided by dressed-state shifts and coherent driving~\cite{Lang11,Hoffman11,Rabl11,PhysRevLett.118.133604,Tang19,PhysRevLett.129.043601,Tang:23}. In contrast, interference-based photon blockade suppresses two-photon excitation through destructive interference between competing excitation pathways, even in weakly nonlinear systems~\cite{Liew10,Bamba2011,Flayac17,Snijders18,Vaneph18,PhysRevA.100.063817,PhysRevLett.127.240402,PhysRevA.90.063824,Tang15,PhysRevA.92.023838,PhysRevA.89.031803}. The interplay between spectral anharmonicity and quantum interference has also been investigated in nonlinear polaritonic systems~\cite{PhysRevLett.125.197402,PhysRevA.101.063824}. A central challenge is to suppress two-photon excitation without weakening the resonant single-photon response, thereby obtaining strong antibunching with a finite cavity population.

Electromagnetically induced transparency (EIT) offers a useful  platform for addressing this question. In a three-level system, destructive interference suppresses absorption through the optically excited state~\cite{Harris1990,Fleischhauer05}. When incorporated into cavity QED, EIT gives rise to dark-state polaritons that combine cavity photons with long-lived atomic coherences~\cite{Wu08,Mucke10,Albert2011}. Their photonic and atomic components can be controlled by coherent coupling fields, allowing the resonance frequencies and optical response to be tuned. 
Cavity EIT has been investigated for enhancing optical nonlinearities and modifying photon correlations through dressed-state interactions~\cite{Rebic_1999,PhysRevA.61.011801,PhysRevA.65.043806,rebic2002photon,Tang21,Tang_2022}. Related approaches have been explored for photon blockade and strongly interacting polaritons in coupled cavities, multilevel emitters, and Rydberg-interacting atomic systems~\cite{Greentree_2000,Hartmann06,Peyronel12,Firstenberg2013,bajcsy2013photon,Chang14,tiarks2019photon}. However, the presence of an EIT-like polariton resonance does not necessarily imply photon blockade. While the dark polariton provides resonant access to the single-excitation manifold, the probability of absorbing a second photon depends on the structure and interference of the two-excitation pathways. Controlling these processes simultaneously is therefore essential for achieving strong antibunching without suppressing the cavity response.

Coherent exchange interactions offer another degree of freedom for manipulating collective light--matter excitations~\cite{anderlini2007controlled,ravets2014coherent,norcia2018cavity}. In systems with two long-lived atomic states, exchange coupling produces symmetric and antisymmetric collective excitations with different energies. Their optical accessibility can be further controlled through the spatial dependence and relative phases of the light--matter coupling~\cite{AsenjoGarcia2017,Ruostekoski2023,PhysRevLett.130.173601,seubert2025tweezer}. These properties have motivated studies of photon blockade and collective quantum-optical responses in multi-emitter cavity-QED systems~\cite{PhysRevA.95.063842,Zhu21,PhysRevA.96.011801,chen2022photon,PhysRevA.99.053850}. 
In cavity EIT, exchange coupling can shift the dark-polariton resonance, but a resonance shift alone does not ensure suppression of the second-photon transition. An important question is whether exchange-induced changes in the one- and two-excitation manifolds can bring the one-polariton resonance into alignment with destructive two-photon interference. The relative phase of the cavity couplings provides an additional means of controlling the participation of symmetric and antisymmetric excitation channels, potentially modifying the photon statistics without changing the exchange interaction.

Here we investigate how coherent spin-exchange interactions and phase-dependent light--matter coupling control photon statistics in a driven ring cavity containing two three-level emitters. By examining the one- and two-excitation manifolds, we identify an EIT-like polariton whose resonance can be tuned relative to the destructive-interference minimum of the second-photon transition amplitude. We show that suitable exchange coupling and control-field strength bring these two conditions into close spectral alignment, producing strong photon antibunching while maintaining a finite intracavity population. We further examine the role of the relative cavity-coupling phase, which redistributes optical coupling between symmetric and antisymmetric collective channels with opposite exchange shifts, allowing the photon statistics to switch between antibunching and bunching without changing the total collective coupling strength. This control of polariton resonances and multiphoton excitation pathways may offer opportunities for tunable nonclassical-light sources and quantum photonic interfaces, including semiconductor quantum-dot platforms coupled to optical microcavities~\cite{Lodahl2015,ma2026recent}.

\section{Model and theoretical framework}

\begin{figure}[ht]
\includegraphics[width=1\columnwidth]{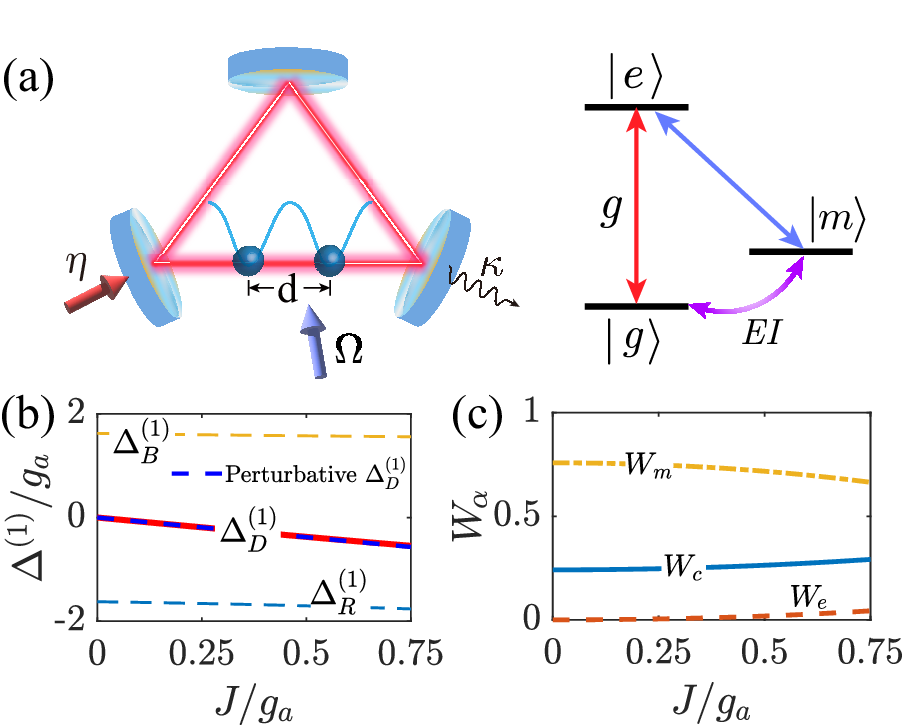}
\caption{\textbf{ System and exchange-tunable cavity-EIT polariton.}
(a) Schematic of two three-level emitters coupled to a single-mode cavity. The $|g\rangle\leftrightarrow|e\rangle$ transition couples to the cavity with strength $g$, the $|e\rangle\leftrightarrow|m\rangle$ transition is driven with Rabi frequency $\Omega$, and the metastable excitations are coupled by the exchange interaction $J$. The cavity is weakly driven with amplitude $\eta$.
(b) One-polariton resonances $\Delta_R^{(1)}$, $\Delta_D^{(1)}$, and $\Delta_B^{(1)}$ versus $J/g$ for the representative value $\Omega/g=0.8$. The blue dashed curve shows the perturbative prediction for $\Delta_D^{(1)}$ from Eq.~(\ref{eq:darkres}).
(c) Cavity ($W_c$), excited-state ($W_e$), and metastable-state ($W_m$) composition weights of the $D$ polariton, showing its predominantly metastable character together with a finite photonic component and weak excited-state admixture.}
\label{fig:model}
\end{figure}

{\em Model and Hamiltonian.---}
We consider two identical three-level emitters coupled to a single traveling-wave mode of an optical ring cavity, as illustrated in Fig.~\ref{fig:model}(a). The cavity couples to the $|g\rangle\leftrightarrow|e\rangle$ transition of each emitter with equal strength $g$, but with a position-dependent phase $\phi=k(x_2-x_1)$, where $x_j$ denotes the position of emitter $j$ along the cavity mode. We assume that the counterpropagating cavity mode is sufficiently suppressed or detuned to be neglected. The $|e\rangle\leftrightarrow|m\rangle$ transitions are driven by a classical control field with Rabi frequency $\Omega$ and a common phase at the two emitter positions. A coherent exchange interaction of strength $J$ transfers a metastable excitation between the two emitters, while the cavity is weakly driven by a coherent probe of amplitude $\eta$.

In the frame rotating at the probe frequency and setting $\hbar=1$, the Hamiltonian reads
\begin{align}
H={}&\Delta_c a^\dagger a+\Delta_e\sum_{j=1}^{2}\sigma_{ee}^{(j)}+\delta_m\sum_{j=1}^{2}\sigma_{mm}^{(j)}\nonumber\\
&+g\left(a^\dagger\sigma_{ge}^{(1)}+a\sigma_{eg}^{(1)}\right)+g\left(e^{i\phi}a^\dagger\sigma_{ge}^{(2)}+e^{-i\phi}a\sigma_{eg}^{(2)}\right) \nonumber\\
&+\Omega\sum_{j=1}^{2}\left(\sigma_{em}^{(j)}+\sigma_{me}^{(j)}\right)+J\left(\sigma_{gm}^{(1)}\sigma_{mg}^{(2)}+\sigma_{gm}^{(2)}\sigma_{mg}^{(1)}\right) \nonumber\\
&+\eta(a+a^\dagger),
\label{eq:H}
\end{align}
where $a$ ($a^\dagger$) annihilates (creates) a cavity photon, and $\sigma_{\alpha\beta}^{(j)}=(|\alpha\rangle\langle\beta|)_j$ denotes the transition operator of emitter $j$, with $\alpha,\beta\in\{g,e,m\}$. The two long-lived states $|g\rangle$ and $|m\rangle$ form an effective pseudospin-$1/2$ manifold, within which the interaction $J(\sigma_{gm}^{(1)}\sigma_{mg}^{(2)}+\sigma_{gm}^{(2)}\sigma_{mg}^{(1)})$ describes coherent spin exchange between the emitters. Here $J$ denotes the exchange-coupling strength, while $\phi$ is the relative phase of the cavity couplings to the two emitters. These two parameters provide complementary control over the polariton spectrum and excitation pathways. For simplicity, we set $\Delta_e=\delta_m=\Delta_c$.  

The dissipative dynamics are governed by the master equation
\begin{align}
\dot{\rho}={}&-i[H,\rho]+\kappa{\cal D}[a]\rho+\gamma_{ge}\sum_{j=1}^{2}{\cal D}[\sigma_{ge}^{(j)}]\rho \nonumber\\
&+\gamma_{me}\sum_{j=1}^{2}{\cal D}[\sigma_{me}^{(j)}]\rho+\gamma_m\sum_{j=1}^{2}{\cal D}[\sigma_{gm}^{(j)}]\rho.
\label{eq:master}
\end{align}
where ${\cal D}[O]\rho=O\rho O^\dagger-\{O^\dagger O,\rho\}/2$.  The decay rates $\gamma_{ge}$, $\gamma_{me}$, and $\gamma_m$ correspond to the transitions $|e\rangle\to|g\rangle$, $|e\rangle\to|m\rangle$, and $|m\rangle\to|g\rangle$, respectively. The steady-state photon statistics are characterized by $g^{(2)}(0)=\langle a^\dagger a^\dagger aa\rangle_{\rm ss}/\langle a^\dagger a\rangle_{\rm ss}^{2}$ and $n_s=\langle a^\dagger a\rangle_{\rm ss}$.  We evaluate the steady-state observables using the full master equation and employ a non-Hermitian weak-drive analysis to identify the coherent excitation pathways.  We first examine exchange-controlled polariton-resonance matching and photon blockade at $\phi=0$, and then investigate how the relative cavity-coupling phase modifies the photon statistics at fixed $J$. 

{\em Cavity-EIT polariton and one-photon resonance .---}
For $\phi=0$, the two emitters couple in phase to the cavity, which therefore accesses only the symmetric collective states $|E_+\rangle=\frac{|eg\rangle+|ge\rangle}{\sqrt{2}},$ $|M_+\rangle=\frac{|mg\rangle+|gm\rangle}{\sqrt{2}}.$
Neglecting the weak probe, the one-excitation Hamiltonian in the basis $\{|1,gg\rangle,|0,E_+\rangle,|0,M_+\rangle\}$ is
\begin{equation}
H_1=
\begin{pmatrix}
\Delta_c & \sqrt{2}g & 0\\
\sqrt{2}g & \Delta_e & \Omega\\
0 & \Omega & \delta_m+J
\end{pmatrix}.
\label{eq:H1}
\end{equation}

For  $J=0$, the cavity and control-field couplings generate an exact dark polariton under the common-detuning condition $\Delta_e=\delta_m=\Delta_c$,
\begin{equation}
|D\rangle=\frac{\Omega|1,gg\rangle-\sqrt{2}g|0,M_+\rangle}{\sqrt{\Omega^2+2g^2}},
\label{eq:darkpolariton}
\end{equation}
which contains no component of the lossy excited state $|E_+\rangle$.
Within the symmetric one-excitation manifold, the exchange Hamiltonian is $H_J=J|0,M_+\rangle\langle0,M_+|$ and shifts the dark polariton through its metastable component. In first-order perturbation theory for the lossless Hamiltonian, the shift is $\delta E_D=\langle D|H_J|D\rangle=2Jg^2/(\Omega^2+2g^2)$.
Because the common detuning $\Delta_c$ shifts the entire one-excitation manifold uniformly, the corresponding one-polariton resonance is
\begin{equation}
\Delta_{D}^{(1)} \simeq -\delta E_D = -\frac{2Jg^2}{\Omega^2+2g^2}.
\label{eq:darkres}
\end{equation}

To incorporate dissipation, we introduce the non-Hermitian one-excitation Hamiltonian
\begin{align}
\mathcal{H}_1 &= H_1 - \frac{i\kappa}{2}|1,gg\rangle\langle 1,gg|
- \frac{i(\gamma_{ge}+\gamma_{me})}{2}|0,E_+\rangle\langle 0,E_+| \nonumber \\
&\quad - \frac{i\gamma_m}{2}|0,M_+\rangle\langle 0,M_+|.
\label{eq:Heff1}
\end{align}

Under the common-detuning condition, ${\cal H}_1(\Delta_c)=\Delta_c I+{\cal H}_1^{(0)}$, with complex eigenenergies ${\cal E}_{1,\nu}(\Delta_c)=\Delta_c+\lambda_{1,\nu}$, where $\nu=R,D,B$ denotes the red-detuned bright, EIT-like dark, and blue-detuned bright polariton branches, respectively. The corresponding resonance positions are $\Delta_\nu^{(1)}=-\operatorname{Re}\lambda_{1,\nu}$. Figure~\ref{fig:model}(b) shows these resonances for $\Omega/g=0.8$, with the first-order prediction in Eq.~(\ref{eq:darkres}) closely following the exact $D$-polariton resonance as $J$ increases.

To characterize the $D$ polariton, we write its normalized right eigenvector as $|R_{1,D}\rangle=c_c|1,gg\rangle+c_e|0,E_+\rangle+c_m|0,M_+\rangle$, with composition weights $W_c=|c_c|^2$, $W_e=|c_e|^2$, and $W_m=|c_m|^2$. Its linewidth is $\Gamma_D=-2\operatorname{Im}\lambda_{1,D}=\kappa W_c+(\gamma_{ge}+\gamma_{me})W_e+\gamma_m W_m$.  As shown in Fig.~\ref{fig:model}(c),  the $D$ polariton retains a predominantly metastable character, a finite photonic component, and a small excited-state admixture throughout the interaction range considered. The finite $W_c$ ensures cavity accessibility, while the small $W_e$ suppresses excited-state losses, favoring resonant single-photon excitation with reduced dissipation.

{\em Dark-polariton-mediated two-photon interference.---}
\label{sec:interference}
Having identified the cavity-accessible $D$ polariton, we now examine the coherent excitation pathways connecting the one- and two-excitation manifolds. At $\phi=0$, the coherent dynamics is confined to the exchange-symmetric sector, with the two-excitation basis $\{|2,gg\rangle,|1,E_+\rangle,|1,M_+\rangle,|0,ee\rangle,|0,EM_+\rangle,|0,mm\rangle\}$, where $|EM_+\rangle=(|em\rangle+|me\rangle)/\sqrt{2}$. The corresponding Hamiltonian $H_2$ and its non-Hermitian counterpart ${\cal H}_2$ are given in Appendix~A. In this manifold, the exchange interaction shifts $|1,M_+\rangle$ by $J$, while the other basis states remain unshifted. It therefore modifies the dressed two-excitation spectrum and the associated transition amplitudes.

In the weak-drive regime, we expand the coherent no-jump state in powers of the probe amplitude $\eta$,
\begin{equation}
|\psi\rangle=|G\rangle+|\psi_1\rangle+|\psi_2\rangle+\cdots,\qquad |G\rangle=|0,gg\rangle,
\label{eq:wavefunction}
\end{equation}
where $|\psi_n\rangle=O(\eta^n)$ denotes the amplitude in the $n$-excitation manifold. To leading order, the stationary amplitude equations yield
\begin{equation}
|\psi_1\rangle=-\eta{\cal H}_1^{-1}a^\dagger|G\rangle,\qquad |\psi_2\rangle=-\eta{\cal H}_2^{-1}a^\dagger|\psi_1\rangle.
\label{eq:psi12}
\end{equation}
The non-Hermitian Hamiltonians account for dissipation through complex energies, while quantum-jump recycling is excluded from this coherent amplitude expansion.

We introduce the biorthogonal right and left eigenstates of ${\cal H}_n$,
\begin{equation}
{\cal H}_n|R_{n,\mu}\rangle={\cal E}_{n,\mu}|R_{n,\mu}\rangle,\qquad
\langle L_{n,\mu}|{\cal H}_n={\cal E}_{n,\mu}\langle L_{n,\mu}|,
\end{equation}
with $\langle L_{n,\mu}|R_{n,\nu}\rangle=\delta_{\mu\nu}$. Near the EIT-like resonance $\Delta_D^{(1)}$, the $D$-polariton contribution dominates the one-excitation response, allowing the single-pole approximation
\begin{equation}
{\cal H}_1^{-1}\simeq
\frac{|R_{1,D}\rangle\langle L_{1,D}|}{{\cal E}_{1,D}}.
\label{eq:darkpole}
\end{equation}
At resonance, $\operatorname{Re}{\cal E}_{1,D}=0$, while the finite imaginary part determines the polariton linewidth. The first photon therefore resonantly excites the $D$ polariton.

Substituting Eq.~(\ref{eq:darkpole}) into Eq.~(\ref{eq:psi12}), we obtain the two-cavity-photon amplitude
\begin{equation}
C_{2c}\equiv\langle2,gg|\psi_2\rangle\simeq\eta^2\frac{\langle L_{1,D}|a^\dagger|G\rangle}{{\cal E}_{1,D}}{\cal M}_D,
\label{eq:C2factor}
\end{equation}
where ${\cal M}_D=\langle2,gg|{\cal H}_2^{-1}a^\dagger|R_{1,D}\rangle$ describes the coherent second-photon transition from the selected $D$ polariton to the two-cavity-photon state. Equation~(\ref{eq:C2factor}) separates resonant one-polariton excitation, governed by $1/{\cal E}_{1,D}$, from the subsequent second-photon transition, governed by ${\cal M}_D$.

To resolve the second-photon excitation pathways, we expand the two-excitation resolvent in its dressed-state basis,
\begin{equation}
{\cal H}_2^{-1}=\sum_\mu\frac{|R_{2,\mu}\rangle\langle L_{2,\mu}|}{{\cal E}_{2,\mu}},
\label{eq:H2spectral}
\end{equation}
which gives
\begin{equation}
{\cal M}_D=\sum_\mu{\cal A}_{\mu D},\qquad {\cal A}_{\mu D}=\frac{\langle2,gg|R_{2,\mu}\rangle\langle L_{2,\mu}|a^\dagger|R_{1,D}\rangle}{{\cal E}_{2,\mu}}.
\label{eq:AD}
\end{equation}
Each ${\cal A}_{\mu D}$ represents a coherent pathway through the $\mu$th two-excitation dressed state. The complex energies and transition matrix elements determine the relative amplitudes and phases of these pathways, allowing the second-photon transition to be suppressed through destructive interference.

To distinguish coherent cancellation from an overall reduction of the pathway amplitudes, we define the interference ratio
\begin{equation}
{\cal R}_D=\frac{\left|\sum_\mu{\cal A}_{\mu D}\right|}{\sum_\mu|{\cal A}_{\mu D}|}=\frac{|{\cal M}_D|}{\sum_\mu|{\cal A}_{\mu D}|},\qquad 0\leq{\cal R}_D\leq1.
\label{eq:RD}
\end{equation}
Here ${\cal R}_D=1$ corresponds to constructive addition, whereas ${\cal R}_D\ll1$ indicates strong destructive interference. A small $|{\cal M}_D|$ accompanied by a small ${\cal R}_D$ and finite individual pathway amplitudes identifies coherent cancellation as the origin of second-photon suppression. With finite dissipation, the cancellation generally produces a pronounced minimum rather than an exact zero.
 
 These results suggest that strong photon blockade can arise when resonant excitation of the cavity-accessible $D$ polariton coincides approximately with destructive interference in the subsequent second-photon transition. We refer to this condition as polariton-resonance--interference matching. By adjusting the control-field strength $\Omega$, favorable matching can be maintained over a range of exchange couplings rather than at a single isolated point. Because the $D$ polariton has a finite linewidth, the one-polariton resonance and the interference minimum need not occur at exactly the same detuning. The matching mechanism is examined below by comparing the coherent pathway analysis with photon correlations calculated from the full Lindblad master equation.

{\em Experimental feasibility.---}
The proposed system could be implemented using two individually trapped $^{87}$Rb atoms coupled to a high-finesse optical cavity. Strong single-atom cavity coupling, cavity electromagnetically induced transparency (EIT), and collective coupling of deterministically positioned atomic arrays have been demonstrated experimentally~\cite{gehr2010cavity,Mucke10,PhysRevLett.130.173601}. Two hyperfine ground states can encode the long-lived states $|g\rangle$ and $|m\rangle$, while an optically excited state provides $|e\rangle$, forming the required $\Lambda$ configuration. Optical-tweezer positioning of atoms within a cavity and position-dependent collective scattering have also been demonstrated, providing experimental means to control the relative atom--cavity coupling phase~\cite{seubert2025tweezer,Yan2023}. The control-field Rabi frequency $\Omega$ can be adjusted through the classical driving field, whereas the effective exchange coupling $J$ between the long-lived states could be engineered through Rydberg-assisted interactions. Coherent exchange between individually trapped Rydberg atoms and strong Rydberg-dressed interactions between atomic spins have been demonstrated experimentally~\cite{ravets2014coherent,Jau2016}.

Independent control of $J$ and $\phi$ requires the exchange strength to remain fixed as the cavity-coupling phase is varied, which could be achieved by adjusting the interaction-dressing fields. The single-traveling-mode approximation further assumes that the counterpropagating cavity mode is sufficiently suppressed or detuned.

In the numerical calculations, we take $\kappa$ as the energy unit and use $g/\kappa=6$, $\gamma_{ge}/\kappa=5/6$, $\gamma_{me}/\kappa=1/6$, $\gamma_m/\kappa=0.01$, and $\eta/\kappa=0.10$. These parameters represent a strongly coupled cavity-EIT system with a long-lived metastable state. We vary $\Omega$ and $J$ to explore the photon-blockade regime and identify favorable conditions for resonance--interference matching. The spontaneous-emission branching fractions are chosen as representative parameters rather than assigned to a specific atomic transition.

\begin{figure*}[t]
\centering
\includegraphics[width=0.9\textwidth]{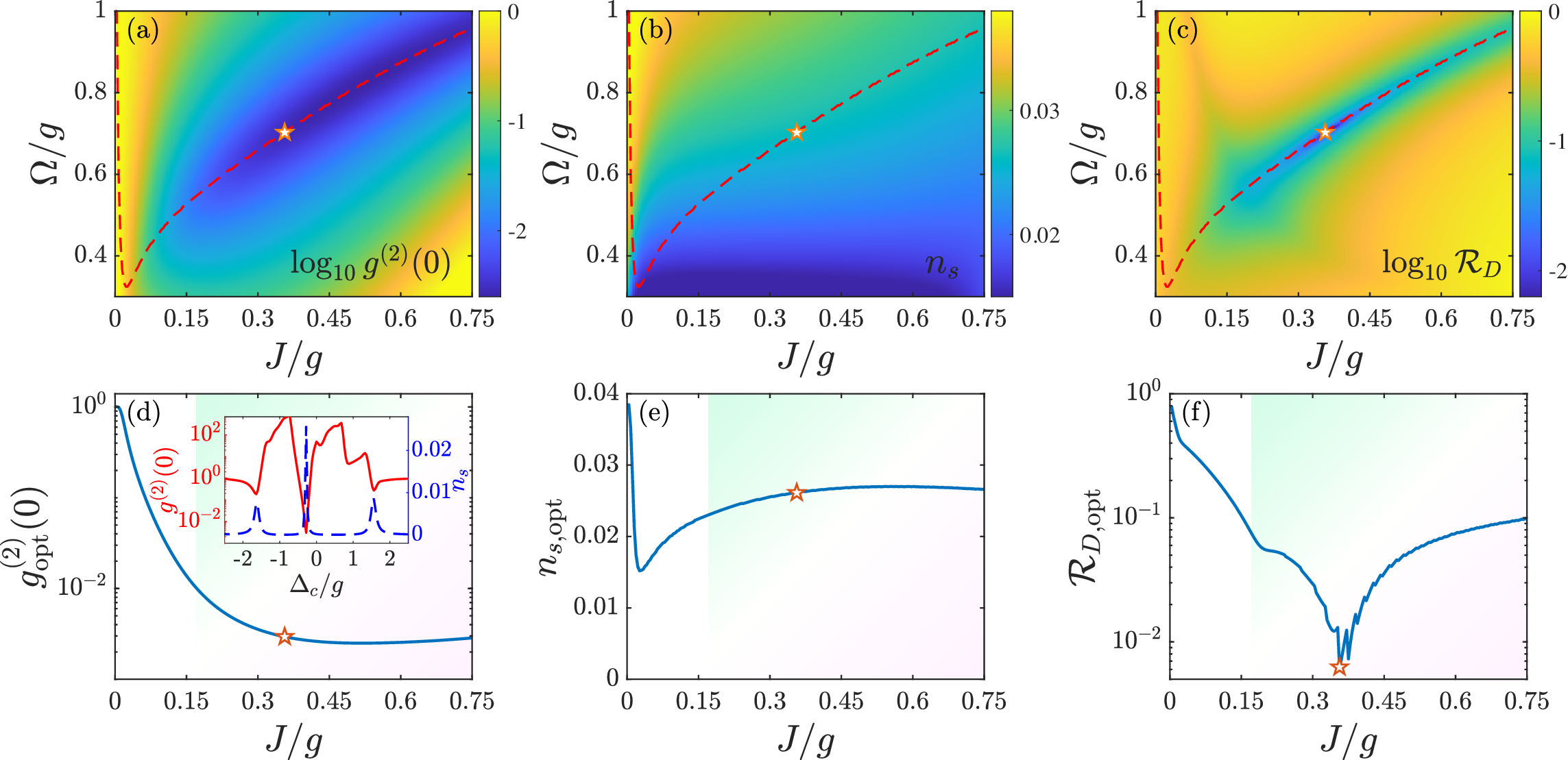}
\caption{
\textbf{Interference-enabled photon blockade.}
(a)--(c) Steady-state photon correlation $\log_{10}g^{(2)}(0)$, intracavity photon number $n_s$, and coherent interference ratio $\log_{10}\mathcal R_D$ as functions of $J/g$ and $\Omega/g$, evaluated along the $D$-polariton resonance $\Delta_c=\Delta_D^{(1)}(J,\Omega)$. The red dashed curves indicate the blockade-optimized trajectory obtained by minimizing $g^{(2)}(0)$ over $\Omega$ for each $J$.
(d)--(f) Corresponding optimal photon correlation $g_{\rm opt}^{(2)}(0)$, cavity population $n_{s,\rm opt}$, and interference ratio $\mathcal R_{D,\rm opt}$ along this trajectory. The stars mark the representative resonance--interference matching point at $J/g=0.35$ and $\Omega/g=0.70$. The inset in (d) shows $g^{(2)}(0)$ and $n_s$ versus $\Delta_c/g$ at this working point.
}
\label{fig:landscape}
\end{figure*}

%%%%%%%%%%%%%%%%%%%%%%%%%%%%%%%%%%%%%%%%%%%%
\section{Results}
\paragraph{Interference-controlled photon blockade.---}
We first examine how the exchange coupling and control field determine the steady-state photon statistics. To maintain resonant access to the first excitation, we evaluate the full Lindblad observables along the EIT-like polariton resonance $\Delta_c=\Delta_D^{(1)}(J,\Omega)$. Figures~\ref{fig:landscape}(a) and \ref{fig:landscape}(b) show the equal-time correlation $g^{(2)}(0)$ and intracavity photon number $n_s$ in the $(J,\Omega)$ plane. A pronounced antibunching valley develops as $J$ increases, but its position changes with $\Omega$. For each $J$, we determine $\Omega_{\rm opt}(J)$ by minimizing $g^{(2)}(0)$ over the plotted control-field range, with $\Delta_c$ constrained to the corresponding one-polariton resonance. The red dashed curves follow this optimized trajectory. Its continuous evolution shows that strong blockade can be maintained at different exchange strengths by adjusting the control field, rather than being restricted to a single isolated operating point.

The correlated dependence on $J$ and $\Omega$ can be understood from their distinct roles in the excitation manifolds. The exchange interaction shifts the metastable component of the $D$ polariton and modifies the two-excitation dressed spectrum, whereas the control field adjusts the polariton composition and the transition amplitudes between excitation manifolds. Tuning $\Omega$ therefore provides a means to maintain favorable resonance and interference conditions as $J$ varies. The extended antibunching valley reflects this joint control rather than the exchange-induced energy shift alone.
 
 Figures~\ref{fig:landscape}(d) and \ref{fig:landscape}(e) show the optimal photon correlation $g_{\rm opt}^{(2)}(0)$ and the corresponding cavity population $n_{s,\rm opt}$ as functions of the exchange coupling. At $J=0$, the cavity field is nearly coherent, with $g_{\rm opt}^{(2)}(0)\simeq1$, indicating that the EIT-like polariton alone does not yield appreciable photon blockade. As $J$ increases, $g_{\rm opt}^{(2)}(0)$ rapidly decreases by more than two orders of magnitude and remains strongly suppressed over an extended interaction range. Importantly, this enhanced antibunching does not result from a vanishing cavity excitation. As shown in Fig.~\ref{fig:landscape}(e), $n_{s,\rm opt}$ initially decreases at weak $J$ but gradually recovers and remains appreciable throughout the strong-blockade regime. This behavior demonstrates that the exchange interaction enables strong suppression of multiphoton excitation while preserving a finite cavity response.

We select $J/g=0.35$ and $\Omega/g=0.70$ as a representative resonance--interference matching point. As shown in the inset of Fig.~\ref{fig:landscape}(d), pronounced photon antibunching occurs near the one-polariton resonance at $\Delta_c/g\simeq-0.28$, where $g^{(2)}(0)\simeq3.0\times10^{-3}$ and $n_s\simeq2.6\times10^{-2}$. 
At this working point, the intracavity population remains approximately $65\%$ of that of an empty resonantly driven cavity under the same drive strength, despite the strong suppression of $g^{(2)}(0)$, demonstrating that strong photon blockade can coexist with a finite intracavity photon population.

\begin{figure*}[t]
\centering
\includegraphics[width=1.9\columnwidth]{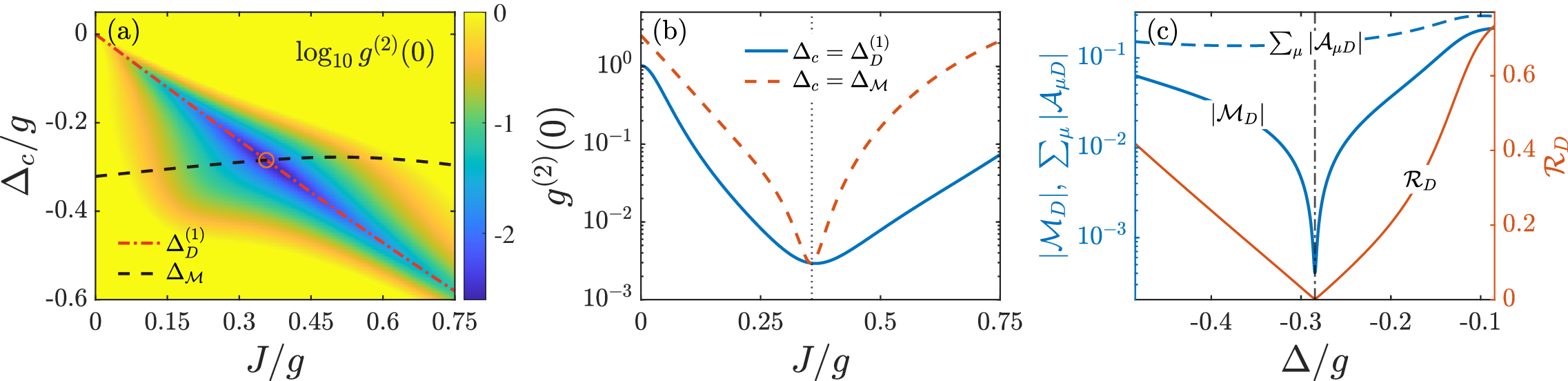}
\caption{\textbf{Polariton-resonance--interference matching and microscopic origin of photon blockade.} (a) $\log_{10}g^{(2)}(0)$ in the $(J/g,\Delta_c/g)$ plane at $\Omega/g=0.70$. The solid and dashed curves denote the EIT-like one-polariton resonance $\Delta_D^{(1)}(J)$ and the coherent interference-minimum trajectory $\Delta_{\mathcal M}(J)$, respectively. The circle identifies the representative matching region. (b) $g^{(2)}(0)$ evaluated along the two independently determined trajectories in (a). (c) Coherent transition-amplitude magnitude $|\mathcal M_D|$, sum of individual dressed-pathway magnitudes $\sum_\mu|\mathcal A_{\mu D}|$, and interference ratio $\mathcal R_D$ versus $\Delta_c/g$ at $J/g=0.35$. The vertical line marks the one-polariton resonance.}
\label{fig:matching}
\end{figure*}

The connection to coherent two-photon interference is revealed by Figs.~\ref{fig:landscape}(c) and \ref{fig:landscape}(f). The low-$\mathcal R_D$ region broadly follows the antibunching valley, indicating that destructive interference among the dressed two-excitation pathways contributes to the extended blockade regime. Along the optimized trajectory, $\mathcal R_{D,\rm opt}$ exhibits a pronounced minimum near the representative matching point, whereas $g_{\rm opt}^{(2)}(0)$ remains strongly suppressed over a wider interaction range. This difference reflects the distinct roles of coherent interference and dissipation: $\mathcal R_D$ quantifies the relative cancellation of the two-photon pathway amplitudes, while the full steady-state $g^{(2)}(0)$ also depends on their absolute strength, the cavity population, and quantum-jump recycling. Consequently, the strongest coherent cancellation need not coincide exactly with the minimum of the observable photon correlation.

\paragraph{Polariton-resonance--interference matching.---}
To identify the microscopic origin of photon blockade, we examine how the exchange interaction controls the one-polariton resonance and the coherent second-photon transition. At fixed $\Omega/g=0.70$, Fig.~\ref{fig:matching}(a) shows $g^{(2)}(0)$ in the $(J,\Delta_c)$ plane. The solid curve denotes the EIT-like one-polariton resonance $\Delta_D^{(1)}(J)$, determined by $\operatorname{Re}\mathcal E_{1,D}=0$, while the dashed curve tracks the interference minimum $\Delta_{\mathcal M}(J)$ obtained from a local minimum of $|\mathcal M_D|$ with respect to $\Delta_c$. Both trajectories are determined independently of the full master-equation photon correlations, allowing a direct comparison between the resonance--interference matching condition and the antibunching region.

The two trajectories exhibit distinct dependences on the exchange coupling. The $D$-polariton resonance redshifts with increasing $J$, consistent with the single-excitation analysis, whereas the interference minimum follows a different dispersion governed by the dressed two-excitation manifold. Remarkably, the two curves nearly coincide at $J/g\simeq0.35$, where $\Delta_D^{(1)}/g\simeq-0.279$ and $\Delta_{\mathcal M}/g\simeq-0.285$. Their separation is much smaller than the $D$-polariton linewidth, allowing resonant single-photon excitation and destructive two-photon interference to occur at nearly the same probe detuning. This connection is verified in Fig.~\ref{fig:matching}(b), where $g^{(2)}(0)$ is evaluated along the independently determined trajectories $\Delta_c=\Delta_D^{(1)}(J)$ and $\Delta_c=\Delta_{\mathcal M}(J)$. Both exhibit pronounced antibunching near the matching region, linking the spectral alignment to the strong photon blockade observed in the full dissipative dynamics.

To understand the microscopic origin of this suppression, we consider the excitation pathways connecting the one- and two-excitation manifolds. Acting on the $D$ polariton with the photon-creation operator gives $a^\dagger|R_{1,D}\rangle=\sqrt{2}c_c|2,gg\rangle+c_e|1,E_+\rangle+c_m|1,M_+\rangle.$
Thus, absorption of the second photon couples the $D$ polariton to several components of the two-excitation manifold, whose coherent response is governed by $\mathcal H_2^{-1}$. The exchange interaction shifts $|1,M_+\rangle$ relative to the other basis states, thereby modifying the dressed-state energies and the complex transition amplitudes $\mathcal A_{\mu D}$. These contributions can interfere destructively, strongly suppressing the resultant two-cavity-photon amplitude $\mathcal M_D$.

Figure~\ref{fig:matching}(c) provides direct evidence of destructive interference among the dressed two-excitation pathways. Near the matching detuning, the coherent transition amplitude is strongly suppressed to $|\mathcal M_D|\simeq4.06\times10^{-4}$, while the sum of individual pathway magnitudes remains finite at $\sum_\mu|\mathcal A_{\mu D}|\simeq0.144$, yielding an interference ratio $\mathcal R_D\simeq2.82\times10^{-3}$. This strong cancellation demonstrates that the suppression of the second-photon transition arises from destructive interference rather than uniformly weak excitation pathways. Together with results in Fig.~\ref{fig:matching}(b), these findings establish resonance--interference matching as the microscopic mechanism underlying the pronounced antibunching near the representative working point.

%%%%%%%%%%%%%%%%%%%

\paragraph{Phase-controlled photon-statistics switching.---}

Having established exchange-controlled photon blockade at $\phi=0$, we now examine how the relative cavity-coupling phase modifies the photon statistics at fixed interaction strength. The phase provides an additional control by redistributing optical coupling between symmetric and antisymmetric collective excitations. Introducing $|E_\pm\rangle=(|eg\rangle\pm|ge\rangle)/\sqrt{2}$ and $|M_\pm\rangle=(|mg\rangle\pm|gm\rangle)/\sqrt{2}$, the phase-dependent cavity couplings are
\begin{equation}
G_+(\phi)=\sqrt{2}g e^{i\phi/2}\cos\frac{\phi}{2},\qquad G_-(\phi)=-i\sqrt{2}g e^{i\phi/2}\sin\frac{\phi}{2}.
\label{eq:gphase}
\end{equation}
The corresponding metastable collective states have opposite exchange shifts, $+J$ and $-J$. In the single-excitation basis $\{|1,gg\rangle,|0,E_+\rangle,|0,E_-\rangle,|0,M_+\rangle,|0,M_-\rangle\}$, the Hamiltonian reads
\begin{equation}
H_1(\phi)=
\begin{pmatrix}
\Delta_c & G_+ & G_- & 0 & 0\\
G_+^* & \Delta_e & 0 & \Omega & 0\\
G_-^* & 0 & \Delta_e & 0 & \Omega\\
0 & \Omega & 0 & \delta_m+J & 0\\
0 & 0 & \Omega & 0 & \delta_m-J
\end{pmatrix}.
\label{eq:H1phase}
\end{equation}
At $\phi=0$ and $\pi$, the cavity couples exclusively to the symmetric and antisymmetric channels, respectively, whereas both channels participate at intermediate phases. Importantly, the total collective coupling strength remains unchanged, $|G_+(\phi)|^2+|G_-(\phi)|^2=2g^2$, independent of $\phi$. Varying $\phi$ therefore redistributes the optical coupling between two exchange-split channels without reducing its total strength. At $\phi=\pi/2$, the two channels have equal coupling magnitudes, $|G_+|=|G_-|=g$. For the symmetric detuning configuration considered here, the transformation $\phi\rightarrow\pi-\phi$ interchanges their optical weights and gives $\Delta_D^{(1)}(\pi-\phi)=-\Delta_D^{(1)}(\phi)$ for the tracked one-polariton resonance.

\begin{figure}[t]
\centering
\includegraphics[width=\columnwidth]{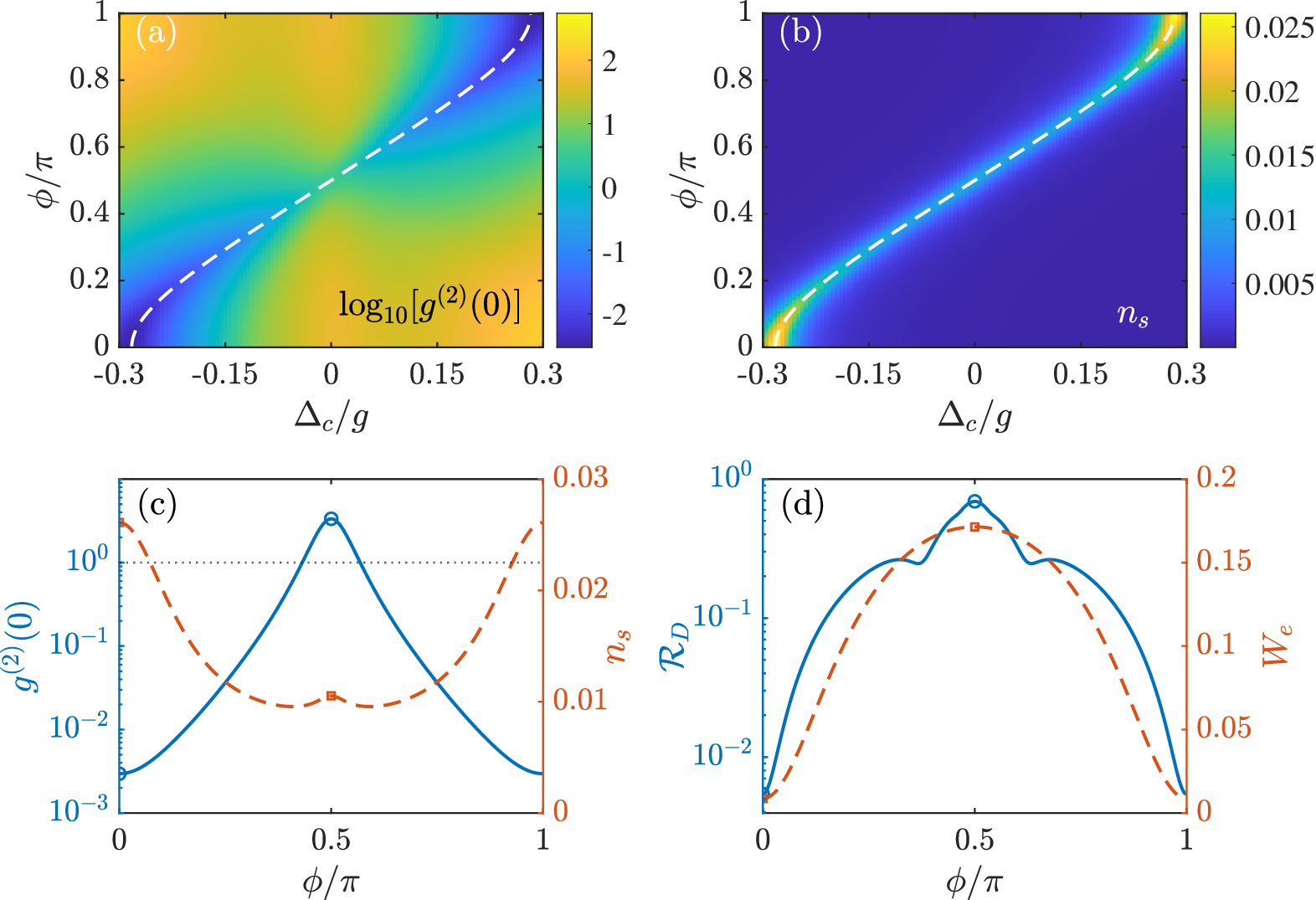}
\caption{\textbf{Phase-controlled photon-statistics switching.} (a) $\log_{10}g^{(2)}(0)$ and (b) $n_s$ in the $(\Delta_c/g,\phi/\pi)$ plane. The white dashed curves indicate the tracked EIT-like one-polariton resonance $\Delta_c=\Delta_D^{(1)}(\phi)$. (c) Photon correlation $g^{(2)}(0)$ and cavity population $n_s$ evaluated along this resonance; the horizontal dotted line marks $g^{(2)}(0)=1$. (d) Coherent second-photon interference ratio $\mathcal R_D$ and excited-state weight $W_e$ of the tracked polariton. The interference ratio is evaluated in the full two-excitation manifold at finite $\phi$. Parameters are $J/g=0.35$ and $\Omega/g=0.70$.}
\label{fig:phase}
\end{figure}

Figures~\ref{fig:phase}(a) and \ref{fig:phase}(b) show the steady-state photon correlation $g^{(2)}(0)$ and intracavity population $n_s$ as functions of $\Delta_c$ and $\phi$, with $J/g=0.35$ and $\Omega/g=0.70$ fixed. The dashed curves follow the EIT-like one-polariton resonance obtained independently from the non-Hermitian single-excitation Hamiltonian. This resonance closely follows the enhanced cavity response in Fig.~\ref{fig:phase}(b), reflecting the finite photonic component of the accessible polariton. In contrast, the photon statistics change markedly along the same resonance, from strong antibunching near $\phi=0$ and $\pi$ to bunching at intermediate phases. Thus, the one-polariton resonance remains optically accessible while the relative coupling phase modifies the multiphoton response.

This behavior is quantified in Fig.~\ref{fig:phase}(c), where both observables are evaluated along the tracked resonance. At $\phi=0$, the cavity exhibits pronounced photon antibunching with $g^{(2)}(0)\simeq3.0\times10^{-3}$. As $\phi$ increases, $g^{(2)}(0)$ rises and reaches approximately $3.35$ near $\phi=\pi/2$, indicating a transition from antibunched to bunched photon statistics. Further increasing $\phi$ restores strong antibunching toward $\phi=\pi$, while the cavity population remains finite throughout the scan. The photon-statistics switching therefore occurs without changing $J$, $\Omega$, or the total collective optical-coupling strength. Moreover, the near-symmetric phase dependence of the photon correlation and cavity population accompanies the reflection of the resonance detuning under $\phi\rightarrow\pi-\phi$.

To elucidate the microscopic origin of this switching, Fig.~\ref{fig:phase}(d) shows the coherent interference ratio $\mathcal R_D$ and the excited-state weight $W_e$ of the tracked polariton, where $W_e=|c_{E_+}|^2+|c_{E_-}|^2$. Near $\phi=0$ and $\pi$, the cavity couples selectively to a single collective channel, and the accessible polariton retains a predominantly metastable character with a small excited-state admixture. The corresponding small $\mathcal R_D$ indicates strong destructive interference among the dressed two-photon pathways. As $\phi$ approaches $\pi/2$, both exchange-split channels contribute with comparable optical strengths, modifying the relative amplitudes and phases of the second-photon transitions. Consequently, $\mathcal R_D$ increases from approximately $5.4\times10^{-3}$ to $0.69$, indicating a substantial weakening of destructive interference, consistent with the photon bunching observed in Fig.~\ref{fig:phase}(c). Meanwhile, $W_e$ increases from approximately $8.7\times10^{-3}$ to $0.17$, reflecting a greater admixture of the lossy excited states and a reduction of the dark-polariton character. These results show that phase-controlled photon-statistics switching is accompanied by changes in both the coherent two-photon interference and the composition of the accessible polariton.

The exchange interaction and the coupling phase thus provide complementary controls of the cavity response. While tuning $J$ and $\Omega$ enables resonance--interference matching over an extended parameter range, varying $\phi$ redistributes optical coupling between exchange-split collective channels, allowing reversible switching between antibunched and bunched photon statistics at fixed $J$ and $\Omega$.

%%%%%%%%%%%%%%%%%%%%%%%

\section{Conclusion}
We have demonstrated spin-exchange-controlled photon blockade in a cavity-EIT system containing two three-level emitters. The coherent spin-exchange interaction affects the one- and two-excitation manifolds differently, shifting the cavity-accessible EIT-like polariton resonance while modifying the dressed two-photon excitation pathways. Strong antibunching occurs when the one-polariton resonance nearly coincides with a destructive-interference minimum of the second-photon transition amplitude. By jointly tuning the exchange coupling and control-field strength, favorable resonance--interference matching can be maintained over an extended parameter range, yielding pronounced photon blockade with a finite intracavity population. 

We further show that the relative cavity-coupling phase provides an additional means of controlling photon statistics. By redistributing optical coupling between symmetric and antisymmetric collective channels with opposite exchange shifts, the phase enables reversible switching between antibunched and bunched photon statistics at fixed interaction strength, without changing the total collective coupling strength. These results establish resonance--interference matching as a mechanism for suppressing multiphoton excitation while preserving resonant single-photon access, without relying primarily on large spectral anharmonicity. Extending this approach to spatially controlled emitter arrays may provide further opportunities to engineer collective excitation pathways and nonclassical photon correlations~\cite{AsenjoGarcia2017,Ruostekoski2023}.

\begin{acknowledgments}
This work was supported by the National Natural Science Foundation of China (Grant No.12374365, Grant No. 12274473, and Grant No. 12135018), Guangdong Provincial Quantum Science Strategic Initiative (Grants No. GDZX2505001), and Guangdong University of Technology SPOE Seed Foundation (SF2024111504).  
\end{acknowledgments}
\

%\acknowledgement
\appendix
\renewcommand{\thefigure}{S\arabic{figure}}
\setcounter{figure}{0}

\section{Two-excitation Hamiltonian}
\label{app:H2}

We derive the Hamiltonian projected onto the symmetric two-excitation manifold used in the coherent weak-probe analysis. At $\phi=0$, the coherent Hamiltonian preserves emitter-exchange symmetry, allowing the symmetric and antisymmetric excitation sectors to be treated separately. This symmetry reduction applies to the no-jump dynamics, whereas the full Lindblad master equation is solved in the complete atomic Hilbert space because local quantum jumps can couple different symmetry sectors. We choose the ordered basis
${\cal B}_2=\{|2,gg\rangle,|1,E_+\rangle,|1,M_+\rangle,|0,ee\rangle,|0,EM_+\rangle,|0,mm\rangle\},$ with $|E_+\rangle=\frac{|eg\rangle+|ge\rangle}{\sqrt{2}},$ $|M_+\rangle=\frac{|mg\rangle+|gm\rangle}{\sqrt{2}},$ and $|EM_+\rangle=\frac{|em\rangle+|me\rangle}{\sqrt{2}}.$

The diagonal energies follow from the bare detunings and the exchange interaction. The latter is described by
\begin{equation}
H_J=J\left(|gm\rangle\langle mg|+|mg\rangle\langle gm|\right),
\end{equation}
which acts only on atomic configurations containing one ground-state and one metastable-state excitation. In particular,
$H_J|M_+\rangle=J|M_+\rangle, $ and $H_J|ee\rangle=H_J|EM_+\rangle=H_J|mm\rangle=0.$
Thus, within ${\cal B}_2$, only $|1,M_+\rangle$ acquires the exchange shift $J$.

The nonzero cavity-induced matrix elements are
\begin{equation}
\begin{aligned}
\langle1,E_+|H_g|2,gg\rangle&=2g,\\
\langle0,ee|H_g|1,E_+\rangle&=\sqrt{2}g,\\
\langle0,EM_+|H_g|1,M_+\rangle&=g.
\end{aligned}
\end{equation}
The factor $2$ in the first matrix element arises from the combined bosonic enhancement and collective atomic coupling. The control field gives
\begin{equation}
\begin{aligned}
\langle1,M_+|H_\Omega|1,E_+\rangle&=\Omega,\\
\langle0,EM_+|H_\Omega|0,ee\rangle&=\sqrt{2}\Omega,\\
\langle0,mm|H_\Omega|0,EM_+\rangle&=\sqrt{2}\Omega.
\end{aligned}
\end{equation}
Combining these contributions, we obtain the two-excitation Hamiltonian
\begingroup
\setlength{\arraycolsep}{1pt}
\begin{equation}
H_2=
\begin{pmatrix}
2\Delta_c & 2g & 0 & 0 & 0 & 0\\
2g & \Delta_c+\Delta_e & \Omega & \sqrt{2}g & 0 & 0\\
0 & \Omega & \Delta_c+\delta_m+J & 0 & g & 0\\
0 & \sqrt{2}g & 0 & 2\Delta_e & \sqrt{2}\Omega & 0\\
0 & 0 & g & \sqrt{2}\Omega & \Delta_e+\delta_m & \sqrt{2}\Omega\\
0 & 0 & 0 & 0 & \sqrt{2}\Omega & 2\delta_m
\end{pmatrix}.
\label{eq:H2app}
\end{equation}
\endgroup

Under the common-detuning condition $\Delta_e=\delta_m=\Delta_c$, all basis states acquire the same detuning contribution $2\Delta_c$. The Hamiltonian can therefore be written as $H_2(\Delta_c)=2\Delta_c I+H_2^{(0)},$ with
\begingroup
\setlength{\arraycolsep}{1pt}
\begin{equation}
H_2^{(0)}
=
\begin{pmatrix}
0 & 2g & 0 & 0 & 0 & 0\\
2g & 0 & \Omega & \sqrt{2}g & 0 & 0\\
0 & \Omega & J & 0 & g & 0\\
0 & \sqrt{2}g & 0 & 0 & \sqrt{2}\Omega & 0\\
0 & 0 & g & \sqrt{2}\Omega & 0 & \sqrt{2}\Omega\\
0 & 0 & 0 & 0 & \sqrt{2}\Omega & 0
\end{pmatrix}.
\label{eq:H20}
\end{equation}
\endgroup

Cavity and atomic losses are incorporated through the non-Hermitian Hamiltonian
\begin{equation}
{\cal H}_2=H_2-\frac{i}{2}\operatorname{diag}(\Gamma_1,\ldots,\Gamma_6),
\label{eq:Heff2}
\end{equation}
where
\begin{align*}
\Gamma_1&=2\kappa,\\
\Gamma_2&=\kappa+\gamma_{ge}+\gamma_{me},\\
\Gamma_3&=\kappa+\gamma_m,\\
\Gamma_4&=2(\gamma_{ge}+\gamma_{me}),\\
\Gamma_5&=\gamma_{ge}+\gamma_{me}+\gamma_m,\\
\Gamma_6&=2\gamma_m.
\end{align*}
These rates account for the total decay out of each basis state through cavity loss, excited-state spontaneous emission, and metastable-state decay. Along the common-detuning trajectory, we obtain
\begin{equation}
{\cal H}_2(\Delta_c)=2\Delta_c I+{\cal H}_2^{(0)},
\end{equation}
where ${\cal H}_2^{(0)}=H_2^{(0)}-\frac{i}{2}\operatorname{diag}(\Gamma_1,\ldots,\Gamma_6)$. The complex dressed energies consequently take the form
$
{\cal E}_{2,\mu}(\Delta_c)=2\Delta_c+\lambda_{2,\mu},
$
with $\lambda_{2,\mu}$ denoting the complex eigenvalues of ${\cal H}_2^{(0)}$.

The exchange interaction enters $H_2^{(0)}$ nonuniformly, shifting only the $|1,M_+\rangle$ component rather than the entire two-excitation manifold. It thereby modifies the relative dressed-state energies and eigenvectors, as well as the complex transition amplitudes governing the second-photon excitation. This provides the microscopic basis for the destructive multipath interference discussed in the main text.

%
%\bibliography{Phase}

%merlin.mbs apsrev4-1.bst 2010-07-25 4.21a (PWD, AO, DPC) hacked
%Control: key (0)
%Control: author (0) dotless jnrlst
%Control: editor formatted (1) identically to author
%Control: production of article title (0) allowed
%Control: page (1) range
%Control: year (0) verbatim
%Control: production of eprint (0) enabled
%

\end{document}